\documentclass{ws-ijmpc}
\usepackage[super]{cite} 

\newcommand{\med}[1]{\left\langle #1 \right\rangle}

\usepackage{color}

\usepackage{amssymb}
\usepackage{graphicx,psfrag}
\usepackage{dcolumn}
\usepackage{bm}
\usepackage{mathbbol}
\usepackage[T1]{fontenc}

\begin{document}

\markboth{F. Author \& S. Author (authors' names)}
{Instructions for typing manuscripts (paper's title)}

%
\catchline{}{}{}{}{}
%


\title{Teleportation based detection of quantum
critical points using small spin chains}

\author{G. A. P. Ribeiro and Gustavo Rigolin\footnote{rigolin@ufscar.br}}

\address{Departamento de F\'{\i}sica, Universidade Federal de S\~ao Carlos, S\~ao Carlos, SP 13565-905, Brazil}

\maketitle

\begin{history}
\received{Day Month Year}
\revised{Day Month Year}
\end{history}

\begin{abstract}
We show for the models here investigated that the teleportation based quantum critical point (QCP) detectors can properly estimate the locations of the QCPs 
when we are not even close to the thermodynamic limit (infinite spin chains) and when we only have access to finite temperature
data. 
Specifically, by working with spin chains with about 
ten qubits 
and in equilibrium with a thermal reservoir
at temperature $T$, we show that it is possible to locate with an error of only a few percents the correct spots of the QCPs for almost all the models studied here. The spin chains we
investigate are given by the XXZ model with or without an external
longitudinal magnetic field as well as the XX model, the XY model,
and the Ising model, all of them subjected to an external
transverse magnetic field.
\end{abstract}


\section{Introduction}

The study of quantum phase transitions 
(QPTs)\cite{sac99}$\,^-\,$\cite{row10}
using the ideas and concepts
stemming from quantum information theory 
made it clear the presence of short and long range quantum correlations around a quantum critical point 
(QCP).\cite{ost02}$\,^-\,$\cite{ike23d}
Indeed, since a QPT is given by a  
change in a macroscopic system's ground state while we modify its Hamiltonian,\cite{sac99} it is expected that quantum correlations should change qualitatively and quantitatively as we cross a QCP.

A QPT is solely caused by quantum fluctuations, which at the QCP
are strong enough to drastically change the macroscopic properties
of the system's ground state. As such, a QPT is theoretically defined at the absolute zero temperature ($T=0$) and in the thermodynamic limit (infinite spin chains).\cite{sac99}
From the experimental point of view, the system's temperature should be low enough such that thermal fluctuations are unable 
to drive the system to its first excited state. In this way 
the system under investigation is effectively always in its ground state while we change its Hamiltonian throughout the parameter space, eventually reaching the QCP. Also, 
the thermodynamic limit is approximated by working with a 
``macroscopic'' system, where the number of molecules are of the order of the Avogadro number ($\approx 10^{23}$). 

If the system's temperature $T$ is sufficiently high, i.e., 
if it is such that $kT$, where $k$ is Boltzmann's constant, 
is of the order of the energy gap between the system's ground state and its first excited state,
the probability to find the system away from its ground state is
not small anymore. In this scenario the characterization of a QPT
is more involved since the influence of the excited states 
in the measured data cannot be ignored. 

The main goal of this manuscript is to investigate how we can 
manage to extract useful information concerning the location 
of a QCP when we are in the ``worst case scenario'', namely, when
we deal with small systems in equilibrium with a thermal
reservoir at temperature $T\neq 0$. In other words, we want to
determine the minimum size of a spin chain 
and the maximum temperature
that still allow us to obtain reliable information regarding the
correct location of the QCP marking a QPT, which 
is rigorously defined at $T=0$ and when the system is in the 
thermodynamic limit (infinite chains).\footnote{Note that the
results reported here for finite chains and in Refs. \refcite{pav23,pav23b,pav24} for infinite chains apply to 
the Hamiltonians (spin chain models) studied here and in those works. There is no general proof yet that the teleportation 
based QCP detectors should work for any model.}

To determine the QCPs of the models we will be
dealing here, we employ one of the most reliable set 
of QCP detectors that provides fairly good estimates of the location of a QCP when only finite $T$ data are available. These
QCP detectors are built on the quantum teleportation 
protocol\cite{ben93,yeo02,rig17} and thus they became known as 
teleportation based QCP detectors.\cite{pav23,pav23b,pav24}
Moreover, these tools were shown to be scalable 
to high dimensional systems (high spin systems), to have a clear experimental meaning, and to work at certain values of $T$ 
where other tools do not.\cite{pav23,pav23b,pav24}
We should also note that there is another recent and 
different tool to detect QCPs at $T = 0$ based on the quantum ``energy'' teleportation protocol.\cite{hot08,hot09} 
It was show in Refs. \refcite{ike23a,ike23b,ike23c,ike23d} 
that for several models the amount of the energy teleported depends on the quantum phase of the system.

This work is organized as follows. In Sec.~\ref{sec2} we review
the main ideas leading to the internal and external teleportation
based QCP detectors, which are 
the main tools used here to detect QCPs.
In Sec.~\ref{sec3} we review the main features associated with 
the spin chain models investigated here. 
In Secs.~\ref{sec4} and \ref{sec5} we present the new results of
this work. We show the usefulness of 
the teleportation based QCP detectors to spot the QCPs for the several models here investigated when we are restricted to deal with small spin chains and with finite $T$ data alone.
In Sec.~\ref{discussion} we discuss more broadly how to use 
these QCP detectors and the requirements for their experimental
implementation. In Sec.~\ref{conclusion} we provide our
concluding remarks.

\section{The teleportation based quantum critical point detectors}
\label{sec2} 

The teleportation based QCP detectors\cite{pav23,pav23b,pav24}
employ two nearest neighbor spins ($\rho_{23}$) 
from a spin chain as the quantum communication channel through which the teleportation protocol\cite{ben93} is implemented. 
It is expected that the macroscopic changes in the 
system's ground state induced by a QPT will also change 
$\rho_{23}$. This will evidently change the efficiency of the teleportation protocol in a very 
perceptible way and thus we can use such intense change in efficiency as a marker of a QPT.\cite{pav23,pav23b,pav24}

The state $\rho_{23}$ 
describing the two nearest neighbor qubits from a 
spin chain as well as the single qubit to be teleported are in general mixed states. As such, we need to 
present the standard teleportation protocol using the 
formalism of density matrices\cite{pav23,pav23b,pav24,rig17,rig15} in order to build the 
teleportation based QCP detectors. 

\begin{figure}
\centerline{\includegraphics[width=5in]{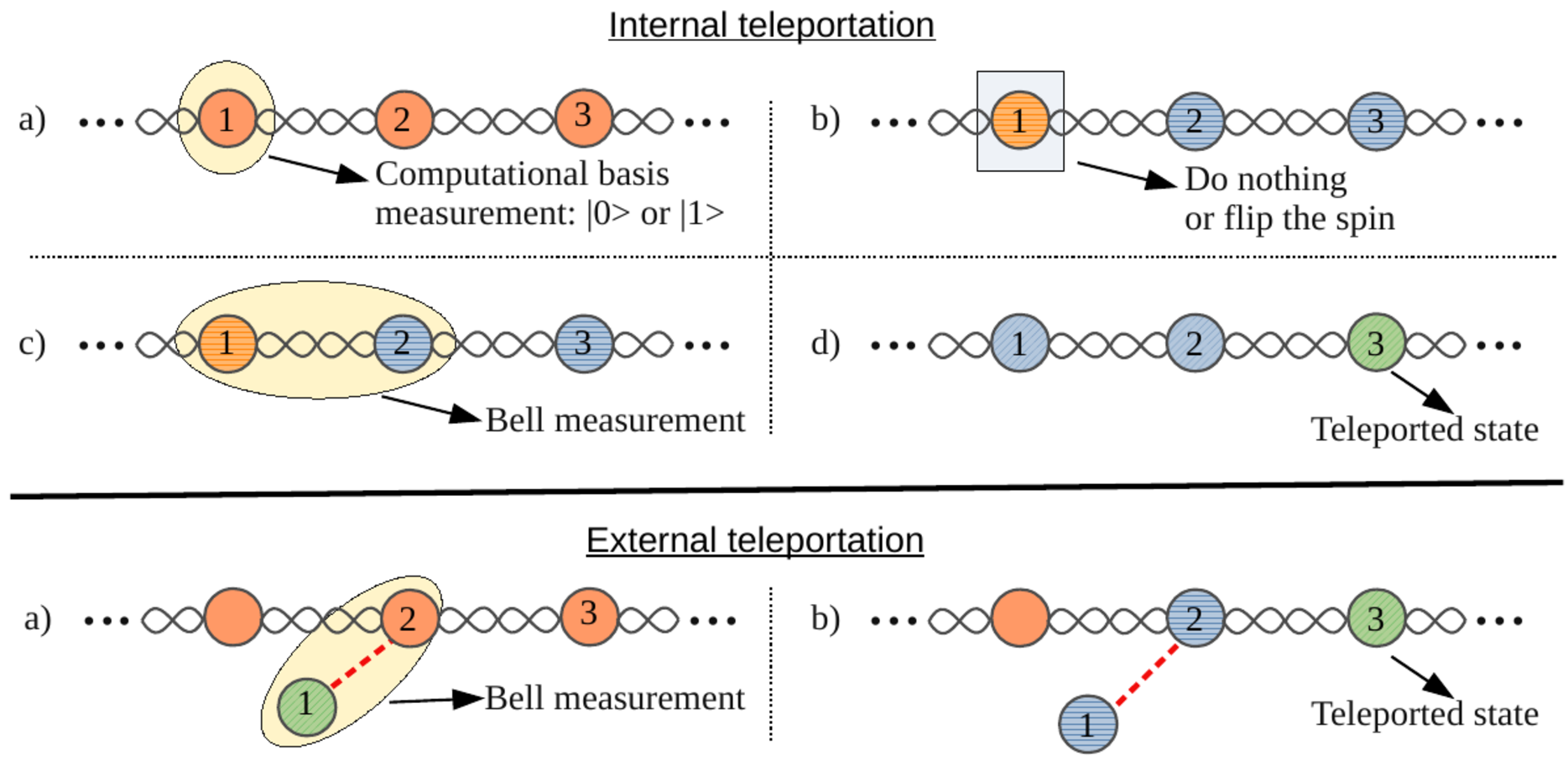}}
\vspace*{8pt}
\caption{Upper (lower) panel: The internal (external)
teleportation protocol. For both cases, Alice and Bob 
choose a pair of spins to be used as the quantum resource to implement the quantum teleportation protocol (qubits $2$ and $3$).
A single run of the internal
teleportation protocol is implemented as follows: 
(a) Alice projects spin $1$ onto the computational basis,
obtaining either the state $|0\rangle$ or $|1\rangle$.
(b) She then either applies onto it the spin flip operation 
($\sigma_1^x$) or does nothing, according to a predetermined
probability. (c) Alice implements a projective
standard Bell measurement (BM) on qubits $1$ and $2$ and 
then informs Bob of her BM result 
via a classical communication channel.
(d) With that information, Bob applies an appropriate unitary operation on his qubit to end the protocol. 
A single run of the external quantum teleportation protocol 
is as follows: Alice brings 
an external qubit to be teleported to Bob (qubit $1$) and
then execute the standard steps of the teleportation protocol
as described in (c) and (d) for the internal teleportation, which
corresponds to (a) and (b) now. See text for more details.} \label{fig0}
\end{figure}

We need three spins (qubits) to fully implement a teleportation
protocol. Qubits $2$ and $3$, given by $\rho_{23}$, 
is the quantum resource shared between Alice (spin $2$) and Bob 
(spin $3$). It is given by tracing out from the spin chain all 
but these two spins. The spin to be teleported (input qubit) 
can either belong to the spin chain (internal teleportation)\cite{pav23b} or be external from it (external teleportation).\cite{pav23} In both situations we call it qubit $1$ 
and it is given by the density matrix $\rho_1$. If the input 
spin belongs to the spin chain, we have the 
internal teleportation based QCP detector, while if it is 
external from the chain, we have the external teleportation 
based QCP detector (see Fig. \ref{fig0}).

For both the internal and external teleportation protocols, 
we want the initial density matrix describing the three qubits before the teleportation starts to be 
given by\cite{pav23,pav23b,pav24} 
\begin{equation}
 \rho = \rho_{1} \otimes \rho_{23},
\label{stepA}
\end{equation}
where $\rho_1$ is the input state and $\rho_{23}$ is the entangled
resource through which the teleportation takes place. 

For the external teleportation, $\rho_1$ is simply the 
external pure state
chosen by Alice while $\rho_{23}$ is the density matrix describing
a pair of qubits from the spin chain. For the internal teleportation, the state describing the three consecutive spins is given by $\rho_{123}$, obtained by tracing out all but qubits 
$1,2$, and $3$ from the state describing the complete spin chain.
This state is not in general given by Eq.~(\ref{stepA}) since 
the three nearest neighbor spins can all be entangled.
Thus, to effectively get Eq.~(\ref{stepA}), 
Alice must implement steps (a) and (b)
described in the upper panel of Fig. \ref{fig0} 
before actually executing the standard teleportation protocol (steps (c) and (d)). 
See Ref. \refcite{pav23b} for the details of how this is done. The 
bottom line is that after those two preparation steps, 
Alice and Bob will share an ensemble of states described 
by Eq.~(\ref{stepA}), where  
$\rho_1$ is the density matrix related to a single spin
of the chain and $\rho_{23}$ is the density matrix associated with
a pair of qubits from the spin chain.

After one run of the quantum teleportation protocol
(see Fig. \ref{fig0}), Bob's spin (qubit 3) is \cite{pav23,pav23b,pav24,rig17}
\begin{equation}
\rho_{_{B_j}}=   \frac{U_jTr_{12}[P_j \rho P_j]U_j^\dagger}{Q_j}.
\label{stepD}
\end{equation}
In Eq.~(\ref{stepD}), $U_j$ represents the unitary 
operation Bob applies onto his qubit. This unitary operation  
depends on which Bell state $j$ Alice obtained in her Bell measurement (BM). The BM projects spins $1$ and $2$ 
onto one of the four Bell states. Note that $Tr_{12}$ means 
the partial trace over Alice's qubits (spins 1 and 2).
Also, $P_j$ is one of the four projectors 
related to a BM,
\begin{eqnarray}
P_{\Psi^{\pm}} &=& |\Psi^{\pm}\rangle \langle \Psi^{\pm}|, \label{projectorA}\\  
P_{\Phi^{\pm}} &=& |\Phi^{\pm}\rangle \langle \Phi^{\pm}|, \label{projectorB}  
\end{eqnarray}
with the Bell states being  
\begin{eqnarray}
|\Psi^{\pm}\rangle&=&(|01\rangle \pm |10\rangle)/\sqrt{2},\label{BellA} \\ 
|\Phi^{\pm}\rangle&=&(|00\rangle \pm |11\rangle)/\sqrt{2}. \label{BellB}
\end{eqnarray}
The quantity 
$Q_j$ in Eq.~(\ref{stepD}) is the probability of Alice 
measuring Bell state $j$,\cite{pav23,rig17}
\begin{equation}
Q_j = Tr[{P_j \rho}].
\label{prob}
\end{equation}

It is important to note that $U_j$ also depends on the quantum state shared between Alice and Bob.\cite{pav23,pav23b,rig17} 
In the standard quantum teleportation protocol, 
where $\rho_{23}$ is a Bell state 
$|k\rangle$, $k=\Psi^{\pm},\Phi^{\pm}$, $U_j$ is chosen from one 
of the following sets $S_k$ of unitary operators,
\begin{eqnarray}
S_{\Phi^+}=\{U_{\Phi^+},U_{\Phi^-},U_{\Psi^+},U_{\Psi^-}\}
=\{\mathbb{1},\sigma^z,\sigma^x,\sigma^z\sigma^x\},
\label{ss1} \\
S_{\Phi^-}=\{U_{\Phi^+},U_{\Phi^-},U_{\Psi^+},U_{\Psi^-}\}
=\{\sigma^z,\mathbb{1},\sigma^z\sigma^x,\sigma^x\}, \label{ss2}\\
S_{\Psi^+}=\{U_{\Phi^+},U_{\Phi^-},U_{\Psi^+},U_{\Psi^-}\}=\{\sigma^x,\sigma^z\sigma^x,\mathbb{1},\sigma^z\}, \label{ss3}\\  
S_{\Psi^-}=\{U_{\Phi^+},U_{\Phi^-},U_{\Psi^+},U_{\Psi^-}\}=\{\sigma^z\sigma^x,\sigma^x,\sigma^z,\mathbb{1}\},\label{ss4}
\end{eqnarray}
in which $\mathbb{1}$ is the identity matrix and $\sigma^\alpha$, 
$\alpha=x,y,z$, is the usual Pauli matrix \cite{nie00}. 
Putting it differently, $S_k$ gives the set of unitary corrections that Bob apply on his qubit if Alice and Bob share the Bell state
$|k\rangle$, with $k=\Psi^\pm,\Phi^\pm$. For example, 
$S_{\Phi^+}$ means that they share the 
state $|\Phi^+\rangle$ and that if her Bell measurement result 
is $|\Phi^+\rangle,|\Phi^-\rangle,|\Psi^+\rangle$, or $|\Psi^-\rangle$, the corresponding unitary operations
that Bob must apply is $\mathbb{1},\sigma^z,\sigma^x$, or $\sigma^z\sigma^x$.

Moreover, since $\rho_{23}$ changes after a QPT, in one quantum 
phase it is best approximated by a specific Bell state while 
in the other phase it is closer to a different one. Hence,
the teleportation based QCP detectors were defined 
by choosing the optimal scenario out of the four sets $S_k$ above.

To quantify the efficiency of the teleportation protocol, we 
use either the fidelity\cite{uhl76,nie00} or the trace distance.\cite{nie00,tos23,tos24} The trace distance will be employed
to quantify the efficiency of the internal teleportation 
protocol since the fidelity has a more complex expression in this case.\cite{pav23b} This happens
because for the internal teleportation, both the input and 
output states are mixed states. Moreover, it can be shown 
that in this scenario the trace distance is also more sensitive to detect QCPs for all the models studied here.\cite{pav23b}
The fidelity, on the other hand, becomes a very simple 
expression for the external teleportation protocol
since the input state
in this case is given by a pure state. Therefore, from now on,
whenever we assess the efficiency of a specific teleportation 
protocol, the fidelity is always associated with the external teleportation while the trace distance to the internal one.

The trace distance between Alice's input qubit and Bob's final state, after a 
single run of the teleportation protocol is \cite{pav23b}
\begin{equation}
D_j(S_k)=D(\rho_1,\rho_{_{B_j}}) = \frac{1}{2}\mbox{Tr}
\left|\rho_1 - \rho_{_{B_j}}\right|,
\label{Dj}
\end{equation}
where $|A| = \sqrt{A^\dagger A}$ and $A^\dagger$ is the adjoint of
$A$.

The geometrical interpretation of the trace distance is such that 
it is half the Euclidean distance between 
the points on the Bloch sphere denoting the two states above.
As such, for two identical states we have $D_j=0$  and 
for two orthogonal pure states we obtain $D_j=1$, the maximum possible value for $D_j$.
For two single qubits we obtain \cite{nie00}
\begin{equation}
D_j(S_k) = \frac{1}{2}\sqrt{(\Delta r_x)^2+(\Delta r_y)^2+(\Delta r_z)^2},
\label{Dj2}
\end{equation}
where $\Delta r_\alpha = \mbox{Tr}(\rho_1\sigma^\alpha)
-\mbox{Tr}(\rho_{_{B_j}}\sigma^\alpha)$. 

After many runs of the teleportation protocol, a given Bell state will be measured with probability $Q_j$. Hence, the relevant quantity in this scenario is the 
mean trace distance,\cite{pav23,pav23b,gor06}
\begin{equation}
\overline{D}(S_k)= \sum_{j=\Psi^{\mp},\Phi^{\mp}} 
Q_jD_j(S_k). \label{Dbar}
\end{equation}
Since the closer two states, the lower the trace distance, 
the optimal choice is obtained by minimizing over all sets $S_k$. Thus, the internal teleportation QCP detector is \cite{pav23b}
\begin{equation}
\overline{\mathcal{D}}_{int} = \min_{\{S_k\}}{\overline{D}(S_k)}.
\label{dmin}
\end{equation}
Equation (\ref{dmin}) is the most accurate QCP detector based on 
the internal teleportation protocol.

For pure input states, the fidelity is
\begin{equation}
F_j(|\psi \rangle,S_k) = \langle \psi | \rho_{_{B_j}} | \psi \rangle,
\label{Fidj}
\end{equation}
where $| \psi \rangle$ is an arbitrary pure state external to the 
chain (see lower panel of Fig.~\ref{fig0}) while the state
$\rho_{_{B_j}}$ is given by Eq.~(\ref{stepD}). Note that the fidelity is maximal (one) if the two states are equal
and minimal (zero) if they are orthogonal. 

Similarly to the trace distance,
the mean fidelity after several runs of the teleportation 
protocol is\cite{pav23,pav23b,gor06}
\begin{equation}
\overline{F}(|\psi\rangle,S_k)= \sum_{j=\Psi^{\mp},\Phi^{\mp}} 
Q_jF_j(| \psi \rangle,S_k). \label{Fbar}
\end{equation}
Now, in addition to maximizing over the four sets $S_k$ 
of unitary operations available to Bob, we also need to maximize
over all pure states $|\psi\rangle$. In this way we obtain the maximum mean fidelity\cite{pav23}
\begin{equation}
\overline{\mathcal{F}}_{ext} = \max_{\{|\psi\rangle,S_k\}}{\overline{F}(|\psi\rangle,S_k)}.
\label{fmax}
\end{equation}
Equation (\ref{fmax}) is our QCP detector stemming from 
the external teleportation protocol. 

\section{The spin chains: general features}
\label{sec3}

The spin chains here studied are all local, i.e., there are 
only nearest neighbor interactions among the spins. Specifically,
we have one dimensional translational invariant spin-1/2 chains, 
with $L$ denoting the number of qubits in the spin chain. 
All models satisfy periodic boundary conditions: 
$\sigma^{x,y,z}_{L+1} = \sigma^{x,y,z}_1$. The subscripts 
in the Pauli matrices label which qubit they act on. 

Before the implementation of the teleportation protocol, the spin chains are in equilibrium with a heat bath at temperature $T$. 
Therefore, the density matrix initially describing the whole chain of $L$ qubits is given by the canonical ensemble density matrix, 
\begin{equation}
\varrho=\frac{e^{-H/kT}}{Z}, \label{canonical} 
\end{equation}
where $Z=Tr[e^{-H/kT}]$ is the partition function and $k$ is 
the Boltzmann's constant.

Tracing out from Eq.~(\ref{canonical}) all but three nearest 
neighbors, we obtain $\rho_{123}$. On the other hand, $\rho_{23}$
and $\rho_{1}$ are obtained from $\varrho$ by tracing out,
respectively, all but two nearest neighbors or all but one spin.
With $\rho_{23}$ and $\rho_{1}$ we can theoretically reconstruct 
Eq.~(\ref{stepA}).

For finite $L$, we numerically compute Eq.~(\ref{canonical}) and
subsequently take the appropriate partial traces as outlined 
in the previous paragraph to obtain $\rho_{12}$ and $\rho_{1}$.
We can also obtain these two density matrices by writing them in terms of one- and two-point correlation functions and then computing those correlation functions
as explained in Refs. \refcite{pav23,pav23b,pav24}.
This later method was the one used by 
Refs. \refcite{pav23,pav23b,pav24} to obtain the infinite spin
chain ($L\rightarrow \infty$) results, which are 
used here as a benchmark for the finite $L$ cases. 
In our numerical analyses for finite $L$, we could go up to 
$12$ spins for the several different values of $kT$ reported in 
the next sections of this work. We also direct the reader to
Refs. \refcite{yan66}-\refcite{zho10}
for all the details related to the calculations
of the one- and two-point correlation functions when 
$L\rightarrow \infty$. See also Ref. \refcite{wer10b}, where 
the relevant results for the calculations of those correlation functions in the thermodynamic limit are reviewed using 
the notation of the present work.

If we abbreviate the one- and two-point correlation functions
as follows,
\begin{eqnarray}
z&=&\med{\sigma_j^z}=\mbox{Tr}[\sigma_j^z\, \varrho], \label{z} \\
ss&=&\med{\sigma_j^s\sigma_{j+1}^s}=
\mbox{Tr}[\sigma_j^s\sigma_{j+1}^s\, \varrho], \label{zz}
\end{eqnarray}
where $s=x,y,z$, we can show that for the models investigated
in the next sections Eqs.~(\ref{dmin}) and (\ref{fmax}) 
become,\cite{pav23,pav23b,pav24}
\begin{eqnarray}
\overline{\mathcal{D}}_{int} &=&
\frac{1}{4}\left[(2 - |z^2 + zz|) |z| + |z^3 - z\cdot zz|\right],
\label{dmin2} \\
\overline{\mathcal{F}}_{ext}&=&
\max{\left[\frac{1 + |xx|}{2}, \frac{1 + |yy|}{2}, 
\frac{1 + |zz|}{2}\right]}.
\label{fmax2a}
\end{eqnarray}
Note that the dot between $z$ and $zz$ denotes the usual multiplication between two real numbers. 

\section{The XXZ models: with and without an external field}
\label{sec4}

The one-dimensional XXZ spin chain is given by the 
following Hamiltonian ($\hbar=1$),
\begin{equation}
H = \sum_{j=1}^{L}\left(\sigma^{x}_{j}\sigma^{x}_{j+1} +
\sigma^{y}_{j}\sigma^{y}_{j+1} + \Delta
\sigma^{z}_{j}\sigma^{z}_{j+1} - \frac{h}{2}\sigma^z_j\right). \label{Hxxz}
\end{equation}
In Eq.~(\ref{Hxxz}) $h$ is an external longitudinal magnetic 
field and the anisotropy $\Delta$ is the tuning parameter.

At zero temperature, this model has two 
QCPs.\cite{yan66}$\,^-\,$\cite{tak99}
For a fixed field $h$, as we vary $\Delta$ we see 
the first QPT at $\Delta=\Delta_1$, where the system's ground
state changes from a ferromagnetic ($\Delta < \Delta_1$) to a critical antiferromagnetic phase ($\Delta_1 < \Delta < \Delta_2$). 
If we keep increasing $\Delta$, we approach another QPT 
at $\Delta=\Delta_2$, after which the system is described by 
an Ising-like antiferromagnet.

When $h=0$ (no external magnetic field), the QCPs
are located at $\Delta_1=-1.0$ and $\Delta_2=1.0$. If we turn 
on the field, the QCPs are displaced. For $h>0$, we have that
the QCPs are displaced to greater values. Without loss of 
generality, in this work we fix the value of the
external field at $h=12.0$. In this scenario, the QCPs are located
at $\Delta_1=2.0$ and $\Delta_2=4.875$, where the latter QCP 
is correct within a numerical error of $\pm 0.001$.\cite{pav23b,pav24}

\begin{figure}
\centerline{\includegraphics[width=5in]{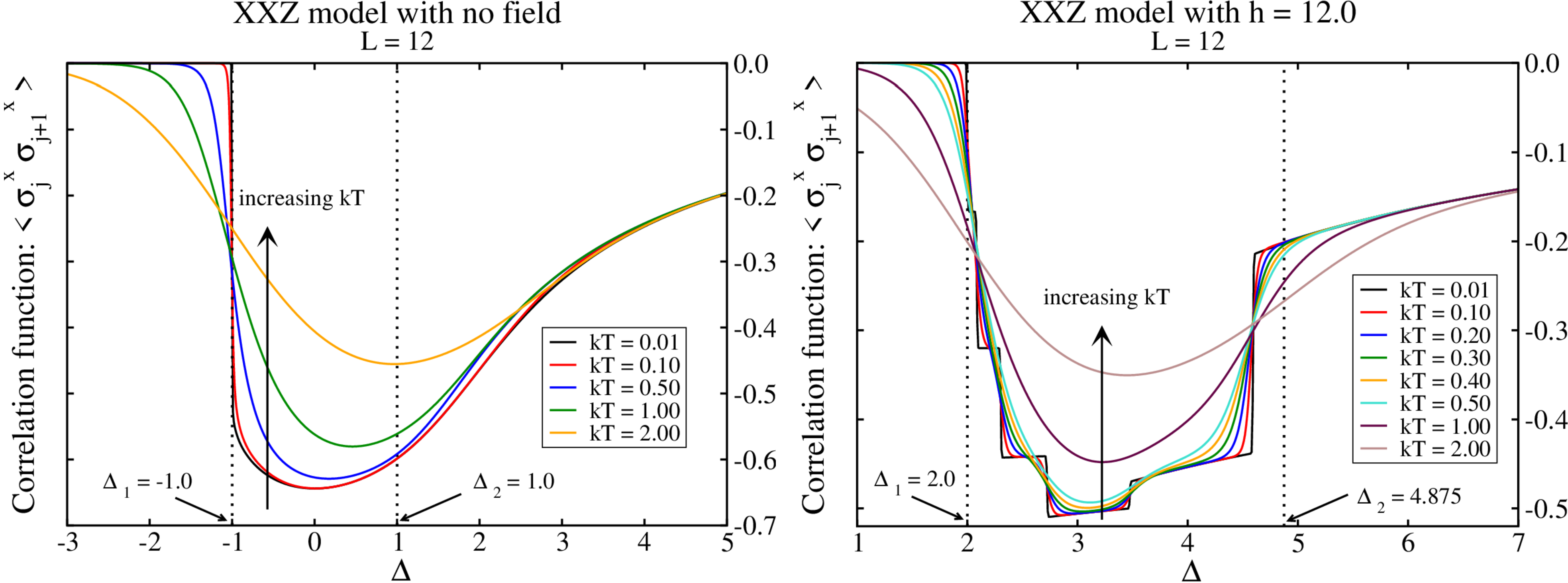}}
\vspace*{8pt}
\caption{The two-point correlation function
$\med{\sigma_j^x\sigma_{j+1}^x}$ for the XXZ model at 
several values of $kT$ as a function of $\Delta$. In the 
left (right) panel, the external field is zero (h=12). 
Here and in all other 
figures all quantities are dimensionless.} \label{fig1}
\end{figure}

Let us start by comparing the behavior of the two-point
correlation function when the system has no field and when it 
has a non zero external field. In Fig. \ref{fig1} we show for $L=12$ the two-point correlation
function $\med{\sigma_j^x\sigma_{j+1}^x}$ for several values
of $kT$. The first point worth noticing is its low $T$ behavior.
When no field is present, $\med{\sigma_j^x\sigma_{j+1}^x}$
is always smooth whenever we are away from the QCPs. On the other
hand, when the external field is turned on, we observe a series
of step-like behavior between the two QCPs. 
The reason for this step-like behavior
can be traced back to the finite size of the system and to 
the fact that when $h\neq 0$ the symmetry of the Hamiltonian 
is reduced, ``lifting'' several degenerate energy levels. The
abrupt changes in the behavior of $\med{\sigma_j^x\sigma_{j+1}^x}$
between the two QCPs are related to the system's ground state being given by
different eigenstates. When $h=0$, the same eigenstate gives the lowest energy
for $\Delta_1<\Delta<\Delta_2$.

Second, as we increase $T$ this step-like behavior diminishes, being absent for sufficiently high $T$. 
This feature is related to the fact that at high 
$T$ the system is described by a mixed state, the canonical ensemble density matrix. In this scenario, several excited 
states, in addition to the ground state, become relevant in 
describing the spin chain, ``washing-out'' the step-like behavior
coming from the ground state when $h\neq 0$.

We should mention that this step-like behavior between the 
QCPs that occurs whenever $h\neq 0$ is also seen in the functional 
form of the other correlation functions or in the behavior of 
other quantities as well. In particular, we will see that 
the fidelity and the trace distance also have a step-like behavior between the QCPs. 
Furthermore, in the 
thermodynamic limit ($L\rightarrow \infty$) 
this step-like feature is absent, even at $T=0$,\cite{pav23,pav23b} since the low-lying excitations 
are gapless.

The important take away from Fig. \ref{fig1} is that
we can discriminate between abrupt changes in the  
correlation functions and in other quantities 
that are related to QCPs from those abrupt changes related to finite size effects. This is achieved by looking at the high $T$ data, 
where we realize that the abrupt changes in the 
correlation functions are no longer present between the two QCPs while near
the two QCPs they can still be seen.

\begin{figure}
\centerline{\includegraphics[width=5in]{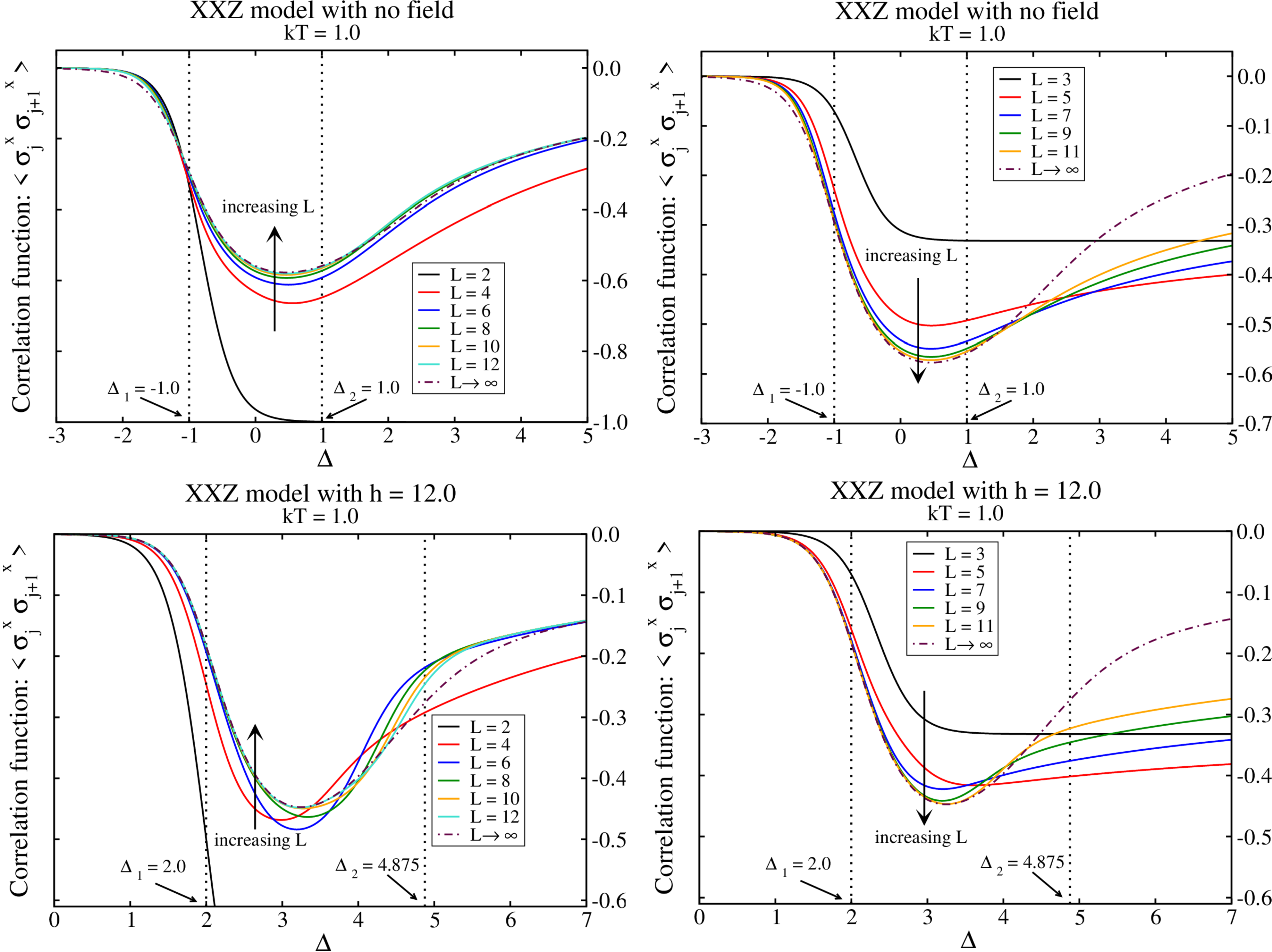}}
\vspace*{8pt}
\caption{The two-point correlation function
$\med{\sigma_j^x\sigma_{j+1}^x}$ for the XXZ model at 
$kT=1.0$ as a function of $\Delta$ for several 
spin chain sizes $L$. In the 
upper (lower) panels, the external field is zero (h=12). 
The left (right) panels show spin chains with an even (odd) 
number of qubits. The dot-dashed curve shows the  
thermodynamics limit case ($L\rightarrow \infty$).} \label{fig2}
\end{figure}

This becomes even clearer if we look at Fig. \ref{fig2}, where
we plot for several values of $L$ the two-point correlation 
function $\med{\sigma_j^x\sigma_{j+1}^x}$ when $kT=1.0$. 
At this temperature, even for small chains, 
of the size of 3 and 4 qubits, we no longer
see the step-like behavior between the two QCPs. Note also that
spin chains with an even number of qubits approach more rapidly
the infinity size case when compared with spin chains with an
odd number of spins. 

Before we move on, we should remark that the above features are also present in the 
other models studied here. As such, from now on we will only deal with spin chains containing an even number of spins and we will always disregard the step-like behavior that
rapidly vanishes as we increase $T$. By proceeding in this way, we will see that very 
good estimates of the correct location of the QCPs can be achieved with spin chains 
having about 10 qubits. 

\begin{figure}
\centerline{\includegraphics[width=5in]{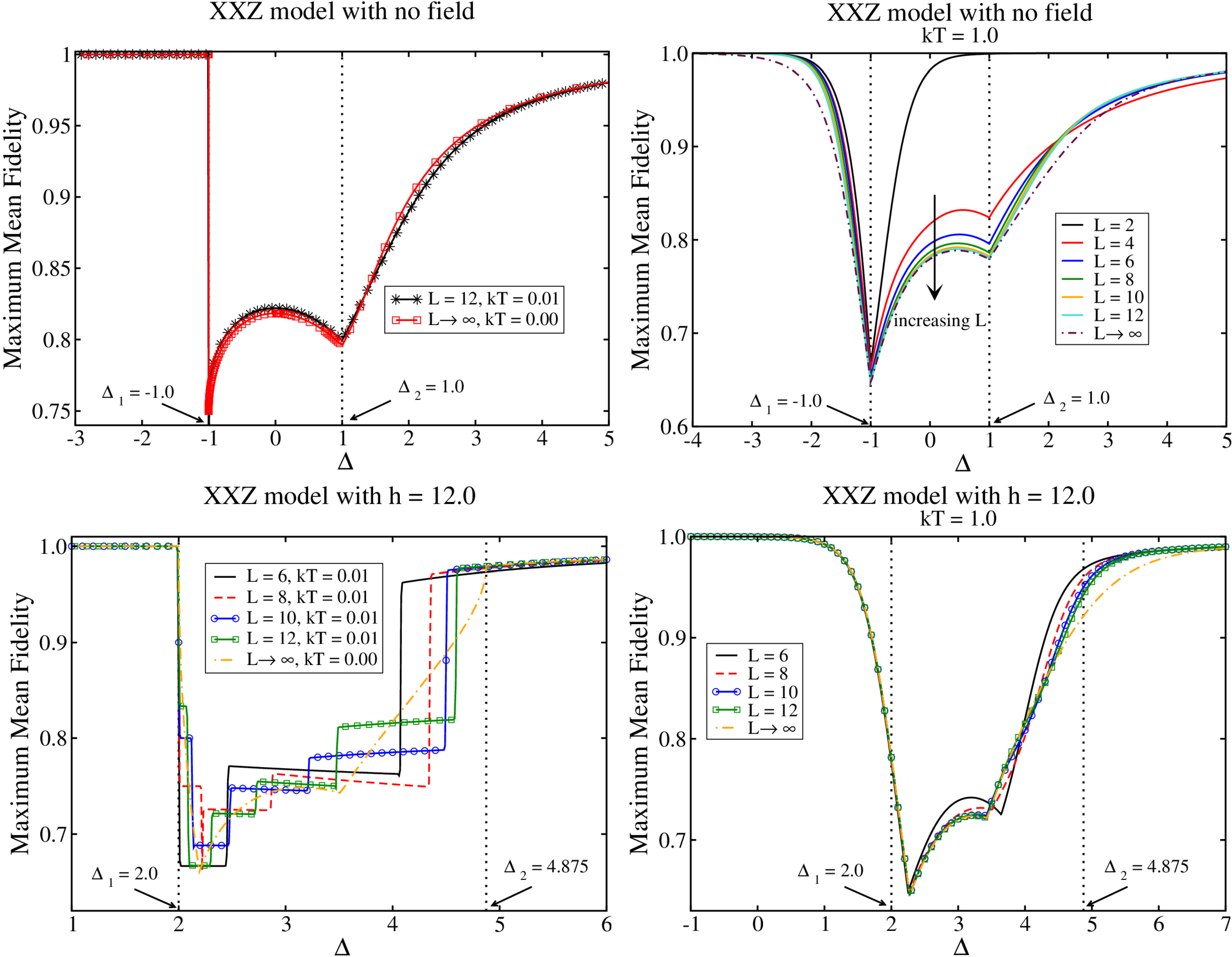}}
\vspace*{8pt}
\caption{Maximal mean fidelity $\overline{\mathcal{F}}_{ext}$ 
[Eq.~(\ref{fmax2a})] as a function of $\Delta$ for the XXZ
model with no field (upper panels) and with an external
magnetic field (lower panels). The left (right) panels show
the low (high) temperature cases.} \label{fig3}
\end{figure}

In the upper panels of Fig. \ref{fig3} we plot 
$\overline{\mathcal{F}}_{ext}$ as given by Eqs.~(\ref{fmax})
and (\ref{fmax2a}) for the case where the external field is zero.
In the upper-left panel, we show the low temperature behavior of
$\overline{\mathcal{F}}_{ext}$ for $L=12$ and 
$L\rightarrow \infty$. Of particular interest is the almost perfect match between both curves. In other words, a spin chain
with about 10 qubits is good enough in providing a fairly accurate approximation of the thermodynamic limit result at $T=0$ when
no external field is present. 
Moreover, at $T=0$ we have that 
$\overline{\mathcal{F}}_{ext}$ is discontinuous at both 
QCPs, clearly indicating the location of those two QCPs.
This feature is also present as we increase $T$ 
(upper-right panel), with $\overline{\mathcal{F}}_{ext}$ having 
discontinuous first order derivatives at the QCPs. 

In the lower panels of Fig. \ref{fig3} we show the maximum mean 
fidelity $\overline{\mathcal{F}}_{ext}$ when the external magnetic
field is turned on ($h=12.0$). As expected, since 
$\overline{\mathcal{F}}_{ext}$ is a function of the correlation
functions [see Eq.~(\ref{fmax2a})], we observe 
the step-like behavior for $\overline{\mathcal{F}}_{ext}$ between
the two QCPs (lower-left panel). As we increase $T$, the 
step-like behavior of $\overline{\mathcal{F}}_{ext}$ 
disappears, with only two cusps remaining between the
two QCPs and two abrupt changes in the value of 
$\overline{\mathcal{F}}_{ext}$ around the locations of the QCPs.
When $T = 0$, these two abrupt changes in the value of 
$\overline{\mathcal{F}}_{ext}$ are given by discontinuities 
in the first order derivatives of $\overline{\mathcal{F}}_{ext}$ 
with respect to $\Delta$. When $T=0$ and $L\rightarrow \infty$, 
both discontinuities in the derivatives occur exactly at the 
locations of the two QCPs and as $T$
increases the cusps (discontinuities in the derivatives)
are displaced from the correct spot of the QCPs and 
smoothed out.\cite{pav23,pav23b}  Also, the two 
other cusps of $\overline{\mathcal{F}}_{ext}$ between the two QCPs 
are related to the optimization over the sets $S_k$ of 
unitary transformations available to Bob.\cite{pav23,pav24}
Before these cusps one set $S_k$ gives the optimal 
$\overline{\mathcal{F}}_{ext}$, while after them another set
$S_k$ is the optimal one.\cite{pav23,pav24}
These cusps are not associated with 
QPTs and they can be distinguished
from the genuine ones marking the two QCPs when $h\neq 0$
by noting that the latter are not related to the
optimization over the sets $S_k$. These cusps 
are actually a consequence of the functional form of 
$\overline{\mathcal{F}}_{ext}$, which depends linearly 
on the magnitude of a particular two-point correlation function 
in the vicinity of the QCP [cf. Eq.~(\ref{fmax2a})].\cite{pav23,pav23b,pav24}

\begin{figure}
\centerline{\includegraphics[width=5in]{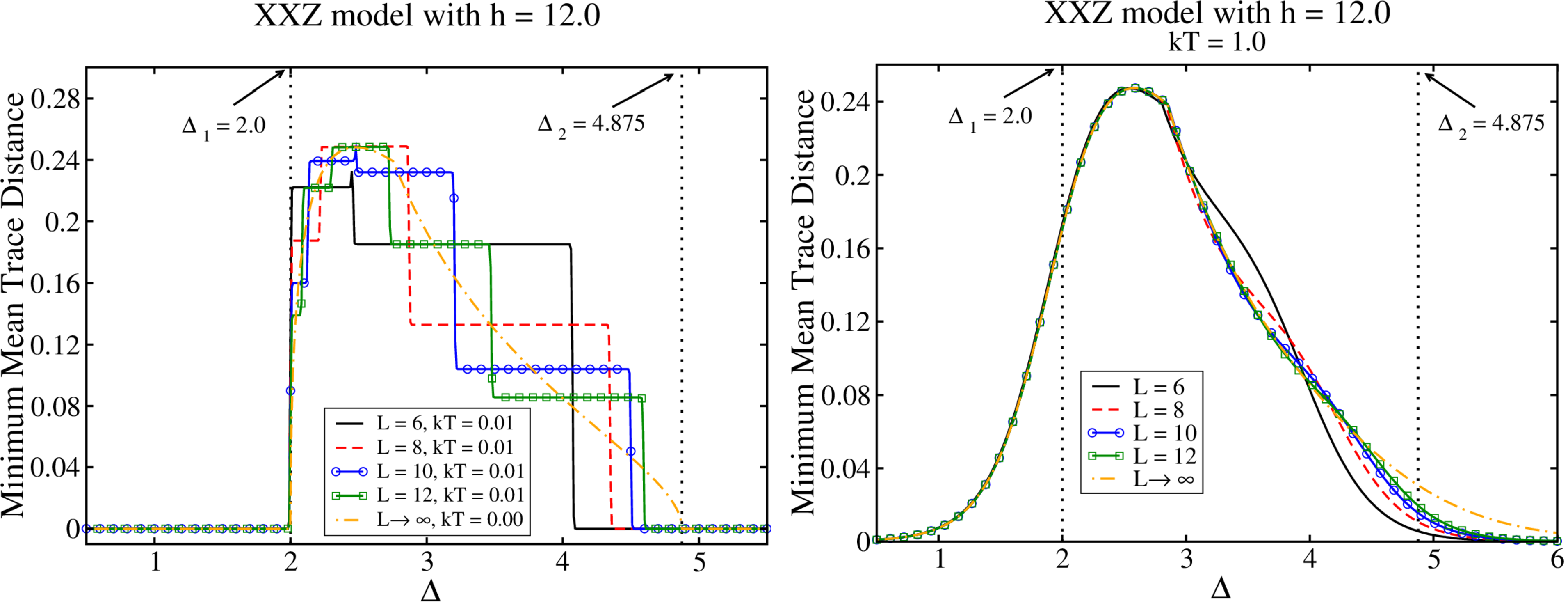}}
\vspace*{8pt}
\caption{Minimal mean trace distance 
$\overline{\mathcal{D}}_{int}$ 
[Eq.~(\ref{dmin2})] as a function of $\Delta$ for the XXZ
model with an external magnetic field. 
The left (right) panel shows the low (high) temperature cases.} \label{fig4}
\end{figure}

Similarly, in Fig. \ref{fig4} we plot 
$\overline{\mathcal{D}}_{int}$
as given by Eqs.~(\ref{dmin}) and (\ref{dmin2}) as a function
of $\Delta$ and when the 
external field is fixed at $h=12.0$. 
We show the curves for several different spin chain sizes
in the low and high $T$ cases. When $h=0$, we always have 
$\med{\sigma_j^z}=0$ and thus 
$\overline{\mathcal{D}}_{int}=0$ for all values of $\Delta$,
illustrating that the internal teleportation is not useful 
to study QPTs in this scenario.\cite{pav23b}

Looking at Fig. \ref{fig4}, we realize that when $T=0$ the QCPs are detected by discontinuities in the derivatives of 
$\overline{\mathcal{D}}_{int}$ with respect to $\Delta$ (see the kinks at $\Delta_1$ and $\Delta_2$ when $L\rightarrow \infty$).  
Now, and contrary to $\overline{\mathcal{F}}_{ext}$, 
we have only one tiny kink between the two QCPs that is not 
related to a QPT. It is located near the first
QCP and just after the maximum of 
$\overline{\mathcal{D}}_{int}$. The origin
of this extra kink is also associated with the optimization over 
the sets $S_k$ of unitary operations that defines 
$\overline{\mathcal{D}}_{int}$\cite{pav23b} and can be distinguished from the genuine 
cusps marking QCPs in the same fashion as described above for the
maximal mean fidelity.\cite{pav23b}

Observing Figs. \ref{fig3} and \ref{fig4}, we note that for 
finite $L$ the first QCP ($\Delta_1$) is very accurately spotted at $T=0$ by discontinuities in the derivatives of either 
$\overline{\mathcal{F}}_{ext}$ or $\overline{\mathcal{D}}_{int}$.
The second QCP ($\Delta_2$), however, is not yet properly 
located with spin chains of the size of up to $L=12$ qubits.
Despite that, if we look at the last cusps of either 
$\overline{\mathcal{F}}_{ext}$ or $\overline{\mathcal{D}}_{int}$,
we see that 
the greater $L$, the closer the cusps are to the exact 
location of
the second QCP (see the left panels of Figs. \ref{fig3} and \ref{fig4}). 
When $T > 0$, the cusps related to the QCPs are smoothed out and displaced from the correct locations of the QCPs. The extra kinks between the QCPs are also displaced from their $T=0$ spots but 
not considerably smoothed out for the temperatures shown in Figs. \ref{fig3} and \ref{fig4}.

To estimate the correct values for the QCPs with finite $T$ data,
we employ the techniques described in Refs. \refcite{wer10b,pav23,pav23b,pav24}. At finite $T$, 
the kinks associated with the QCPs are smoothed out but we 
still see abrupt changes in the 
values of the above quantities around the QCPs. Thus, for a 
given $T$, we calculate the derivatives of 
$\overline{\mathcal{F}}_{ext}$ and $\overline{\mathcal{D}}_{int}$ with respect to $\Delta$ around the QCPs and then select the 
value of $\Delta$ providing the greatest magnitude for the derivatives. This specific $\Delta$ is chosen as the best 
approximation to the location of the QCP at that fixed $T$. 
This procedure is repeated for several temperatures, allowing 
us to extrapolate to $T=0$ and thus obtain a very good approximation to the exact location of the QCP.\footnote{For 
spin chains with 10 or 12 qubits, and taking into account 
the numerical errors to estimate the location of the QCP, 
a linear regression using the finite $T$ data reported here 
is almost always sufficient to identify within a few percents of error ($< 10\%$) the correct location of the QCP  when we take the $T=0$ limit. The numerical error associated with the location of the extremum value of a first (second) order derivative is $\pm0.01 (0.02)$ and it is 
related to the fact that we change $\Delta$ in increments of
$0.01$. See Refs. \refcite{pav23,pav23b,pav24} for more
details.} 

\begin{figure}
\centerline{\includegraphics[width=5in]{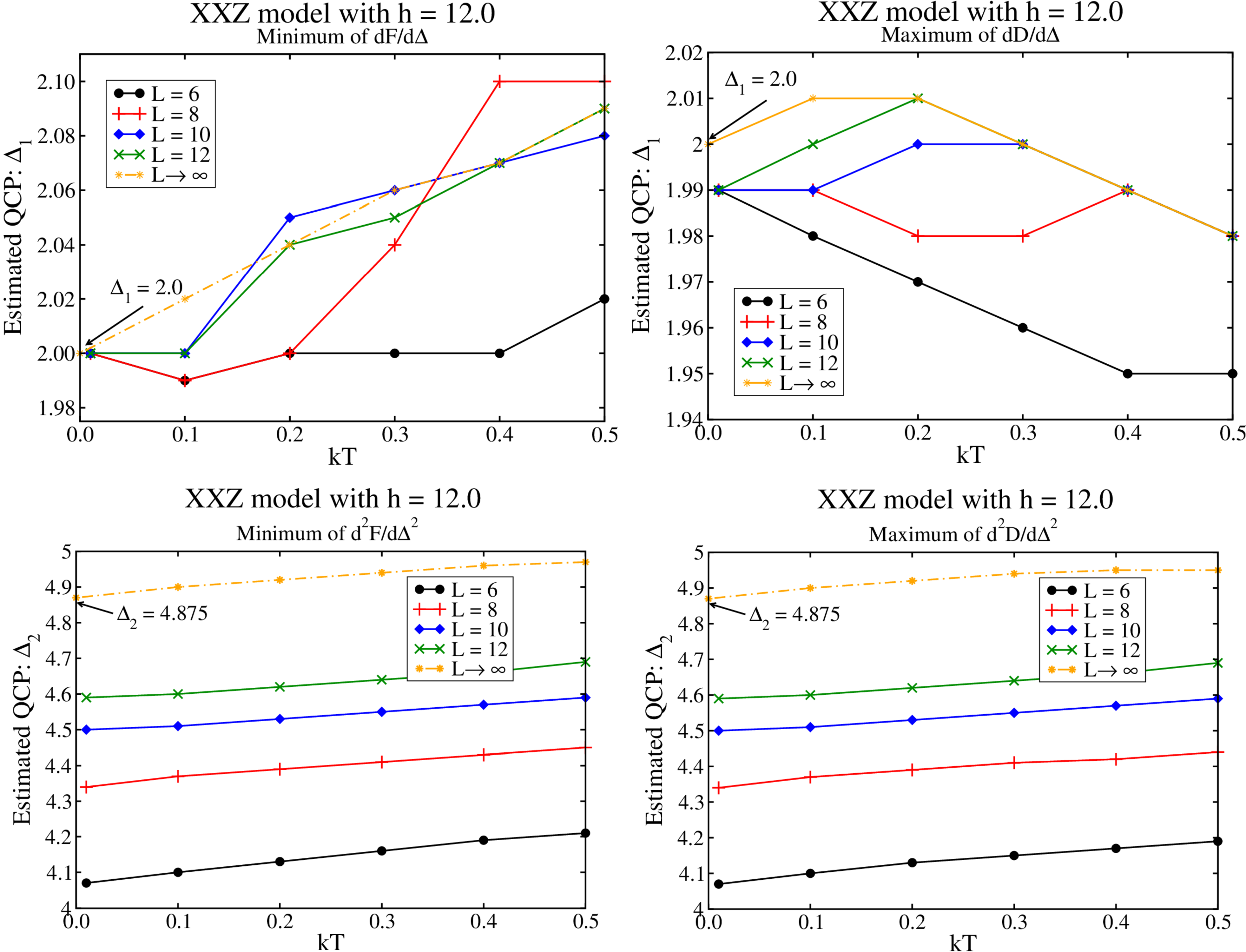}}
\vspace*{8pt}
\caption{Estimated QCPs for the XXZ model in an external field ($h=12.0$) employing finite $T$ data in accord with the recipe detailed in the text. For a fixed $kT$,
the upper panels give the values of $\Delta$ maximizing 
$|d\overline{\mathcal{F}}_{ext}/d\Delta|$ (left) and
$|d\overline{\mathcal{D}}_{int}/d\Delta|$ (right)
and the lower panels give the values of $\Delta$ associated with 
the maximum of $|d^2\overline{\mathcal{F}}_{ext}/d\Delta^2|$
(left) and $|d^2\overline{\mathcal{D}}_{int}/d\Delta^2|$ (right).
Note that the extrema of the derivatives related to the 
finite size step-like behavior of the above quantities 
were disregarded in the current analysis.} \label{fig5}
\end{figure}

We deal with six different values of temperature, 
namely, $kT=0.01, 0.1, 0.2, 0.3,$ $0.4, 0.5$. 
For each $kT$, we calculate $\overline{\mathcal{F}}_{ext}$ and 
$\overline{\mathcal{D}}_{int}$ as a function of $\Delta$ and in increments of $0.01$. Subsequently, around $\Delta_1$ we numerically compute the first order derivatives 
of the above quantities, selecting the value of $\Delta$ giving 
the greatest magnitude for those derivatives. In the vicinity of
$\Delta_2$, we numerically evaluate the second order derivatives of those quantities, selecting again the value of $\Delta$ that 
provides the greatest magnitude for the second order derivatives. 
The values of those $\Delta$'s are shown in Fig. \ref{fig5}
for several chain sizes.

Looking at the upper panels of Fig. \ref{fig5}, we notice that 
the estimates for the QCP $\Delta_1$ are very accurate. For 
small temperatures ($kT \leq 0.1$), we obtain the correct
spot of the QCP with spin chain sizes of the order of $L=10$.
Moreover, using spin chains with $10$ and $12$ qubits we already
get a very good approximation to the results of the thermodynamic limit ($L\rightarrow \infty$). Indeed, taken into account the numerical errors in obtaining the extrema of the first order 
derivatives\cite{pav23,pav23b}, we see that for $L=10$ and $12$
the curves showing the maxima of 
$|d\overline{\mathcal{F}}_{ext}/d\Delta|$ and
$|d\overline{\mathcal{D}}_{int}/d\Delta|$  follow the trend of 
the $L\rightarrow \infty$ case. 

The correct spot of the QCP $\Delta_2$ is not very well estimated
with spin chain sizes of at most $L=12$. However, looking at 
the lower panels of Fig. \ref{fig5}, we realize that for all 
temperatures, the greater $L$, the closer the estimated location of the QCP is to its correct spot. 

\section{The XY and Ising models in an external transverse field}
\label{sec5}

Using the same notation, boundary conditions, and assumptions
of the previous section, the one-dimensional XY model in 
an external transverse magnetic field is described by the
following Hamiltonian,\cite{lie61,bar70,bar71}
\begin{equation}
H \!=\! -\frac{\lambda}{4}\!\sum_{j=1}^{L}\!\left[(1+\gamma)\sigma^{x}_{j}\sigma^{x}_{j+1} + (1-\gamma)\sigma^{y}_{j}\sigma^{y}_{j+1}\right]\! 
- \frac{1}{2}\!\sum_{j=1}^{L}\!\sigma^z_j. \label{Hxy}
\end{equation}
In Eq.~(\ref{Hxy}), $\lambda$ is proportional to the inverse of 
the magnetic field magnitude while $\gamma$ denotes the anisotropy parameter. Fixing $\gamma=\pm 1$ we get the 
transverse Ising model and fixing $\gamma=0$ we obtain the isotropic XX model in a transverse field.

For a fixed $\gamma$, by changing the external field ($\lambda$)
we arrive at a QCP when $\lambda_c=1.0$ (the Ising transition).
When $\lambda < 1$, the ground state is given by an ordered ferromagnet and after crossing the QCP, i.e., when $\lambda > 1$, 
the ground state describes a quantum paramagnet.\cite{pfe70} 
On the other hand, fixing $\lambda$, with $\lambda > 1$, we obtain another QPT by changing $\gamma$ (the anisotropy transition).
The QCP now is located at $\gamma_c=0$  \cite{lie61,bar70,bar71,zho10} and one phase is an ordered ferromagnet in the y-direction and the other one is an ordered ferromagnet in the x-direction. 

As we already remarked in the previous section, the same finite 
size effects reported for the XXZ model show up here. First, 
for low $T$ all relevant quantities show a step-like behavior as we change the tuning parameter. This step-like behavior is not associated with a QPT and it is suppressed as we increase $T$.
Second, spin chains that have an even number of  
qubits approach more rapidly the infinity size case 
than spin chains having an odd number of qubits. Hence,
to save space, we will not show the plots of the correlation
functions, focusing our attention on the study of the 
maximal mean fidelity $\overline{\mathcal{F}}_{ext}$ and 
of the minimal mean trace distance $\overline{\mathcal{D}}_{int}$.

\subsection{The $\lambda$ transitions}

\begin{figure}
\centerline{\includegraphics[width=5in]{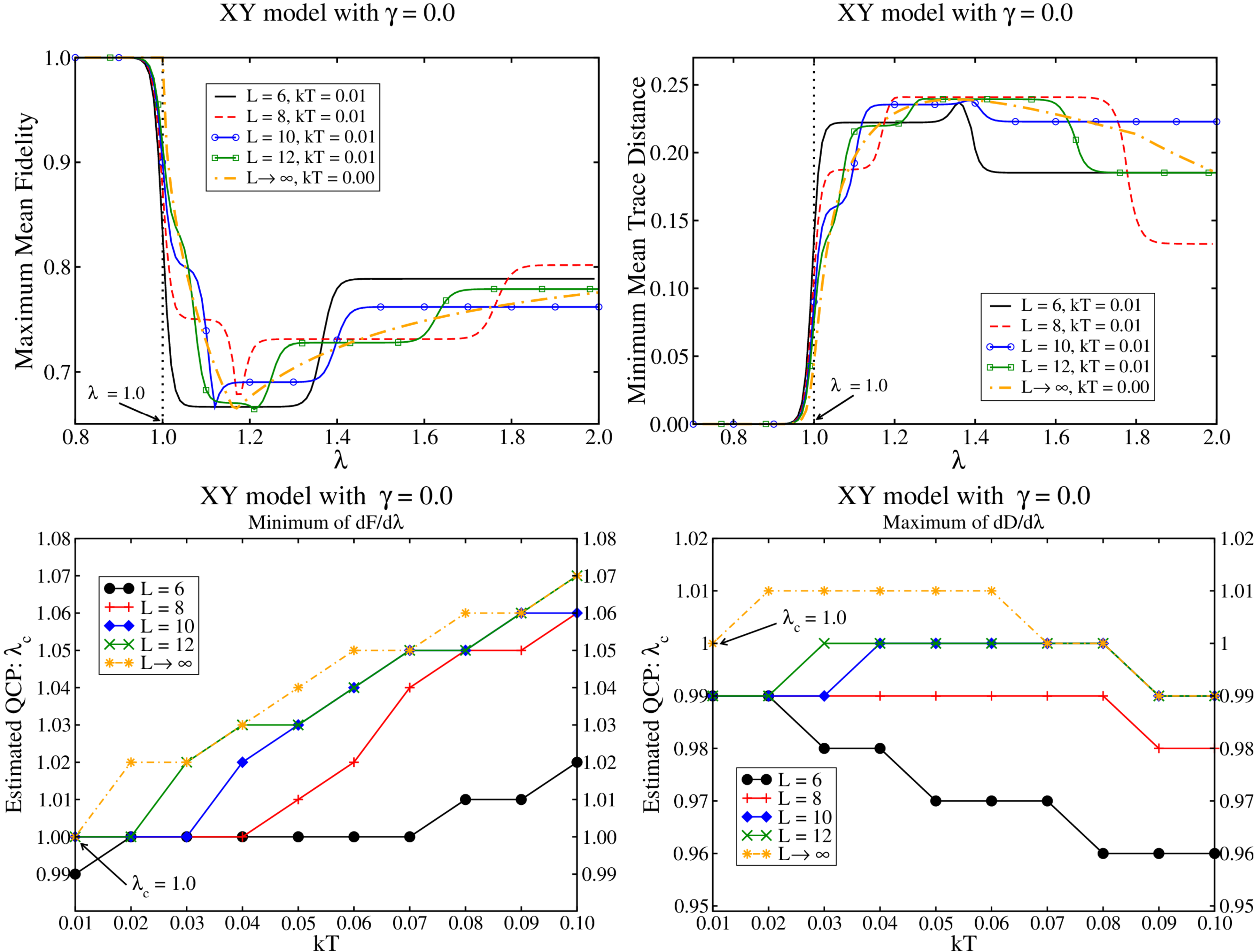}}
\vspace*{8pt}
\caption{The XX model in a transverse field ($\gamma=0.0$).
Upper panels: $\overline{\mathcal{F}}_{ext}$ [left, Eq.~(\ref{fmax2a})] and $\overline{\mathcal{D}}_{int}$ [right, Eq.~(\ref{dmin2})] as functions of $\lambda$ at low $T$.
Lower panels: Estimated QCPs employing finite $T$ data as 
explained in the text. For a fixed $kT$,
the left (right) panel gives the values of $\lambda$ maximizing 
$|d\overline{\mathcal{F}}_{ext}/d\Delta|$ 
($|d\overline{\mathcal{D}}_{int}/d\Delta|$).
We should remark that for the present and following models, 
the extrema of the derivatives coming from the 
step-like behavior related to finite size effects are disregarded.
See text for details.} \label{fig6}
\end{figure}

Starting with the isotropic XX model ($\gamma=0.0$), 
we realize that in the thermodynamic limit the QCP $\lambda_c$ 
is spotted at $T=0$ by discontinuities in the derivatives of 
$\overline{\mathcal{F}}_{ext}$ and $\overline{\mathcal{D}}_{int}$
at $\lambda_c=1.0$ 
(see the upper panels of Fig. \ref{fig6}).\cite{pav23b,pav24}
For $T>0$, the $T=0$ discontinuities in the derivatives manifest 
themselves in very high values for the magnitude of 
those derivatives around the QCP. Moreover, the locations of those extrema move away from the correct spot of the QCP as we increase $T$. Looking at the lower panels of Fig. \ref{fig6}, we recognize
that for $kT \lesssim 0.10$ these extrema are located near the
exact value of the QCP. By properly extrapolating to $kT=0$,
the correct spot of $\lambda_c$ can be predicted. The numerical
analysis leading to the results reported in the lower panels of Fig. \ref{fig6} is exactly the same one already explained in the previous section, when we dealt with the XXZ model in an external field. Finally, it is worth mentioning that by taking into 
account the numerical error to estimate the extrema of the first
order derivatives ($\pm 0.01$), the minimum mean trace distance  
$\overline{\mathcal{D}}_{int}$ pinpoints the correct spot
of the QCP even at the high value range of 
temperature ($kT=0.10$). This happens whenever the spin chain
have at least $10$ qubits.

\begin{figure}
\centerline{\includegraphics[width=5in]{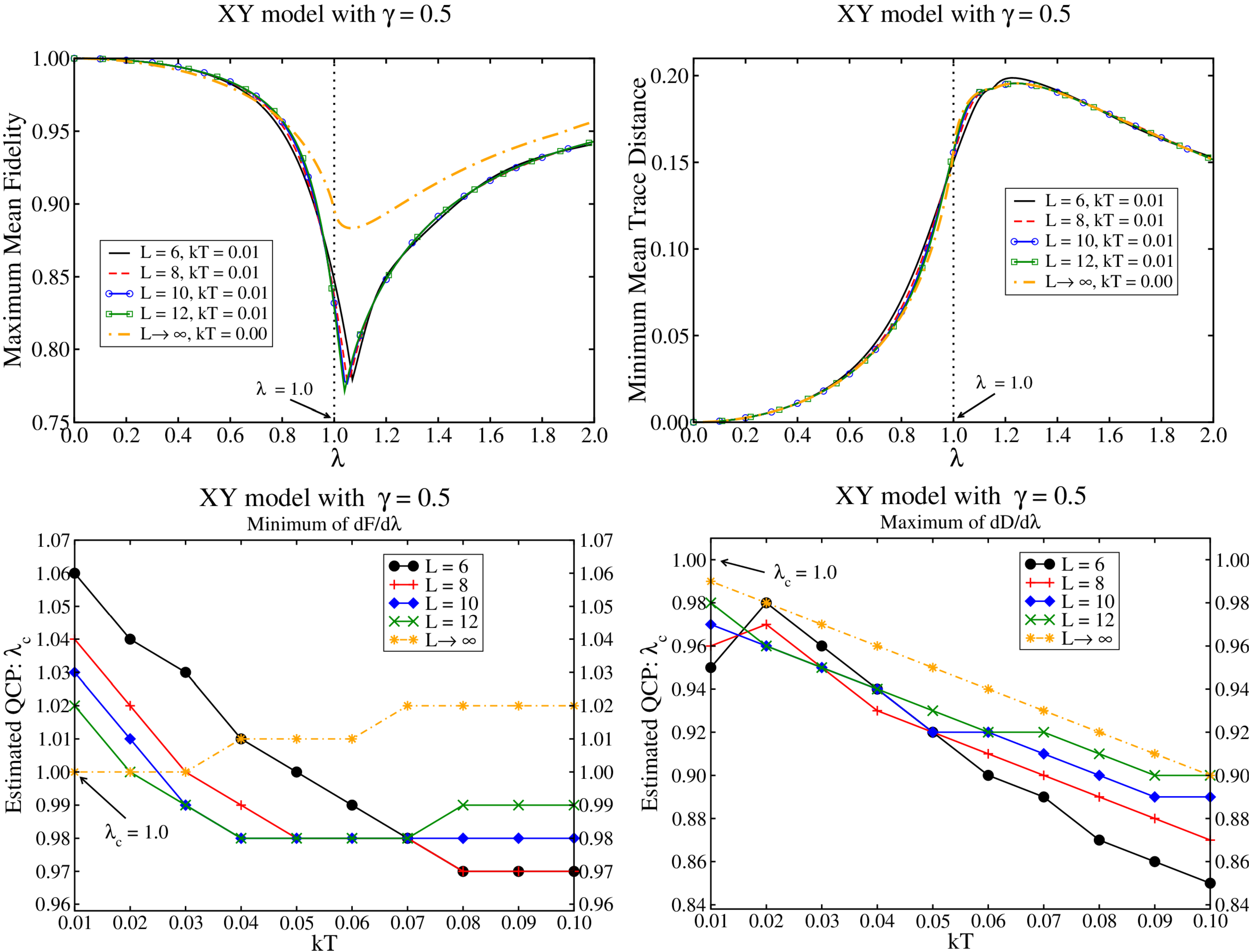}}
\vspace*{8pt}
\caption{The same quantities as explained in Fig. \ref{fig6},
but now we deal with the anisotropic XY model in a 
transverse field ($\gamma=0.5$).} 
\label{fig7}
\end{figure}

When $\gamma=0.5$, namely, when we are dealing with the anisotropic XY model, 
we note that for infinite chains the QCP is given by inflection points of 
$\overline{\mathcal{F}}_{ext}$ and $\overline{\mathcal{D}}_{int}$ 
that occur exactly at $\lambda_c=1.0$ when $T=0$. See the upper panels of 
Fig. \ref{fig7} and Refs. \refcite{pav23b,pav24} for the complete analysis in the  
thermodynamic limit. By increasing $T$, though, the inflection points are displaced 
away from $\lambda_c=1.0$. Proceeding as already explained for the other models, by determining the location of those inflection points for several values of $T$, we can extrapolate to the correct $T=0$ value. 
This is illustrated in the lower panels of Fig. \ref{fig7}, where the extrema of the 
first derivatives of $\overline{\mathcal{F}}_{ext}$ and 
$\overline{\mathcal{D}}_{int}$ 
around the QCP is taken as the best approximation to the latter 
when $T\neq0$.
Looking at the
lower-left panel of Fig. \ref{fig7}, we realize that for finite $L$, the low $T$ 
estimates to the QCP coming from $\overline{\mathcal{F}}_{ext}$ are not too good. 
However, as we increase $L$ we clearly see better and better estimates for the QCP. Interestingly, the estimates for the QCP coming from $\overline{\mathcal{D}}_{int}$
are much better (lower-right panel of Fig. \ref{fig7}), 
with all curves already following the general trend of the 
$L\rightarrow \infty$ case. For $L=12$ we obtain, taking into account the numerical error 
to determine the extrema of $\overline{\mathcal{D}}_{int}$, the same prediction coming from
the thermodynamic limit case at low $T$. Specifically, we get 
$\lambda_c=0.98 \pm 0,01$ for $L=12$ and 
$\lambda_c=0.99 \pm 0.01$ for $L\rightarrow \infty$ when $kT=0.01$. 

\begin{figure}
\centerline{\includegraphics[width=5in]{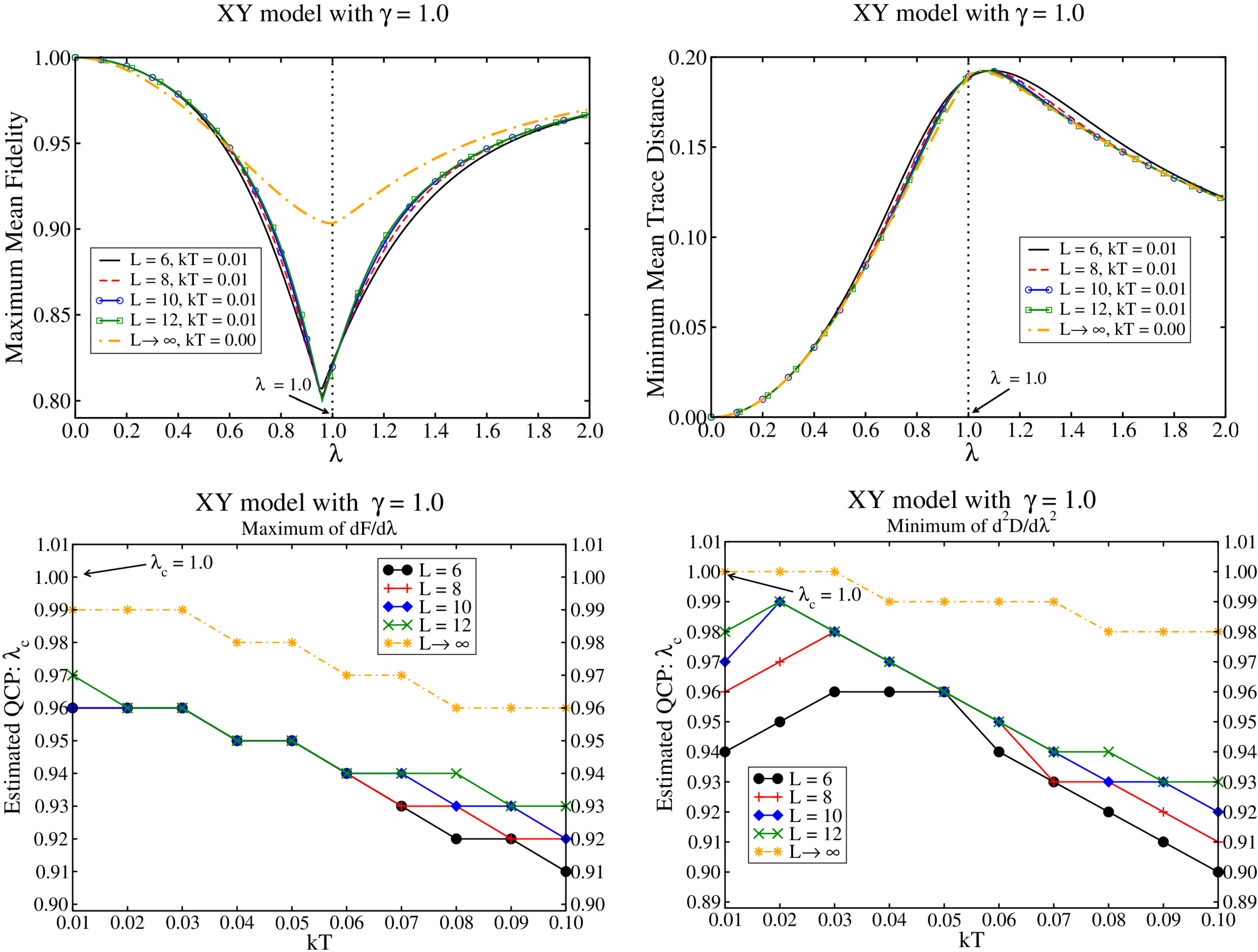}}
\vspace*{8pt}
\caption{The same quantities as explained in Fig. \ref{fig6},
but now we deal with the Ising model in a 
transverse field ($\gamma=1.0$).} 
\label{fig8}
\end{figure}

Setting $\gamma=1.0$, we obtain the Ising transverse model.
In this case, the QCP is determined by an inflection point of 
$d\overline{\mathcal{F}}_{ext}/d\lambda$ and an inflection point 
of $d^2\overline{\mathcal{D}}_{int}/d\lambda^2$, which 
occur exactly at the QCP when $T=0$ and $L\rightarrow \infty$. 
See the upper panels of Fig. \ref{fig8} and in particular 
Refs. \refcite{pav23b,pav24} for more details. In the same 
way as observed for the other models subjected to an external
field, as we increase $T$ the positions of the inflection points
change, being farther from the correct location of the QCP the 
greater the temperature. 
And similarly, by determining the location of those 
inflection points for several temperatures we can extrapolate to 
the correct $T=0$ spot of the QCP. The same analysis already 
applied to the other models leads to the lower panels of 
Fig. \ref{fig8}. Looking at those panels, we note that now the 
location of the inflection points of 
$d\overline{\mathcal{F}}_{ext}/d\lambda$ for finite $L$ 
follow the general trend of the $L\rightarrow \infty$ case. 
The lower $T$, the closer the inflection points are to 
the correct value of the QCP.
At low values of $T$, the estimate of the QCP coming from 
the $L=12$ qubit chain is impressive. In this case we obtain $\lambda_c=0.97\pm0.01$ at $kT=0.01$ against the estimate $\lambda_c=0.99\pm0.01$ when
$L\rightarrow \infty$. The positions of the inflection points
of $d^2\overline{\mathcal{D}}_{int}/d\lambda^2$ behave in a 
similar way. Their high $T$ locations are not too good in 
estimating the correct spot of the QCP while the low $T$ data
are very accurate. When $L=12$, we get $\lambda_c=0.98\pm0.02$ 
at $kT=0.01$ against $\lambda_c=1.00\pm0.02$ when 
$L\rightarrow \infty$. Note that the numerical error in
determining the extrema of the second derivatives is twice
the error in determining the extrema for the first 
derivatives.\cite{pav23b,pav24}

Before we finish this subsection, we should remark that the cusps
seen for $\overline{\mathcal{F}}_{ext}$ in 
Figs. \ref{fig6}-\ref{fig8} close to the QCP are not associated with a QPT. Those cusps are associated with the point where 
$|\langle\sigma_j^x\sigma_{j+1}^x\rangle| = 
|\langle\sigma_j^z\sigma_{j+1}^z\rangle|$. 
Before the cusp $|\langle\sigma_j^x\sigma_{j+1}^x\rangle| < 
|\langle\sigma_j^z\sigma_{j+1}^z\rangle|$ and after it
$|\langle\sigma_j^x\sigma_{j+1}^x\rangle| > 
|\langle\sigma_j^z\sigma_{j+1}^z\rangle|$. This fact together
with the functional form of $\overline{\mathcal{F}}_{ext}$ 
[cf. Eq.~(\ref{fmax2a})] are the causes of those cusps and 
they can be traced back to the optimization over all sets of unitary correction $S_k$ available to Bob in obtaining the final expression for the maximal mean fidelity.\cite{pav23}
A similar argument explains the tiny cusp after the QCP seen for 
the curve of $\overline{\mathcal{D}}_{int}$ when $\gamma=0.0$,
located about $\lambda=1.8$. This cusp can only be seen at the thermodynamic limit (dotted-dashed curve in the upper-right
panel of Fig. \ref{fig6}) and is better depicted in Ref. \refcite{pav23b}. Its cause is the minimization over the sets $S_k$ that leads to 
$\overline{\mathcal{D}}_{int}$ (see Ref. \refcite{pav23b}).
For $\gamma=0.5$, the reason for the spurious 
small cusp seen in the upper-right panel of Fig. \ref{fig7} 
after the QCP is related to the last term in the expression for $\overline{\mathcal{D}}_{int}$ 
[cf. Eq.~(\ref{dmin2})].\cite{pav23b}  
That cusp is the point where $z^3 - z \cdot zz$ changes sign. 
The $\gamma=1.0$ case has no spurious cusps since in this case
$z^3 - z \cdot zz\leq 0$ for all the range of $\lambda$ plotted in
Fig. \ref{fig8} (see Ref. \refcite{pav23b}).

\subsection{The $\gamma$ transition}

\begin{figure}
\centerline{\includegraphics[width=5in]{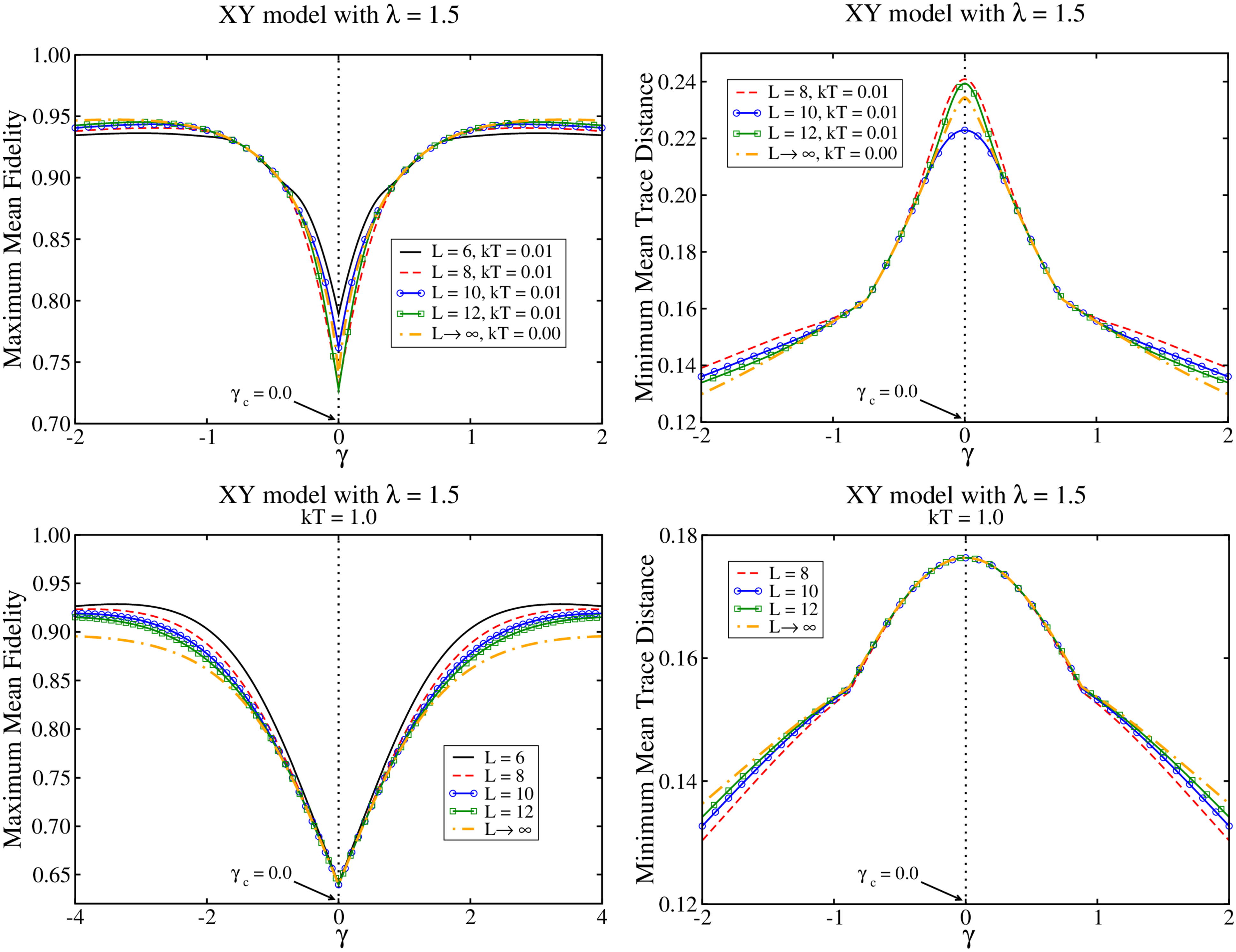}}
\vspace*{8pt}
\caption{Maximal mean fidelity $\overline{\mathcal{F}}_{ext}$
(left panels) and minimal mean trace distance   
$\overline{\mathcal{D}}_{int}$ (right panels) at the low (upper panels) and high (lower panels) temperature ranges.} 
\label{fig9}
\end{figure}

Fixing $\lambda=1.5$, in Fig. \ref{fig9} we show 
$\overline{\mathcal{F}}_{ext}$ and $\overline{\mathcal{D}}_{int}$
as functions of $\gamma$ for several chain sizes $L$. 
The QCP $\gamma_c=0.0$ is given by a minimum of $\overline{\mathcal{F}}_{ext}$, located 
exactly at $\gamma_c=0.0$ for all the values of $T$ shown in
Fig. \ref{fig9}. Note that this minimum is also a cusp, a point
where the derivative of $\overline{\mathcal{F}}_{ext}$ is
discontinuous, and it does not move away from $\gamma_c$ as we 
increase $T$. On the other hand, a global maximum of 
$\overline{\mathcal{D}}_{int}$ spots the exact location of 
the QCP $\gamma_c$. This maximum of 
$\overline{\mathcal{D}}_{int}$ is not displaced as $T$ increases
as well.
Note that the two cusps away from the QCP in the curves of
$\overline{\mathcal{D}}_{int}$ are not associated with QPTs. 
They occur when $z^3 - z \cdot zz$ changes sign, 
which leads to a discontinuity
in the derivative of $\overline{\mathcal{D}}_{int}$ since the 
latter depends on the absolute value of 
$z^3 - z \cdot zz$ [cf. Eq.~(\ref{dmin2})].\cite{pav23b}

\section{Discussion}
\label{discussion}

The main goal of this work was a systematic investigation of 
how we can determine the correct spot of a QCP  
when we are not in the thermodynamic limit (infinite spin chains).
In particular, we showed for several
models if the teleportation based QCP detectors,\cite{pav23,pav23b,pav24} one of the most reliable 
classes of QCP detectors, can provide
good estimates for the location of the QCPs when we are restricted
to work with spin chains containing about 10 qubits.
The results reported here support the view that it is indeed 
possible to infer the correct spot of the QCP using spin chains
containing just a few tens of qubits. Moreover, these results 
should be understood more broadly and under two different 
yet complementary points of view. 

The first one 
assumes that we do know the thermodynamic limit behavior for 
the model under investigation. In this scenario, we have a
complete characterization of the system's phase diagram and 
thus a clear benchmark to compare the finite size analysis. 
This comparison allows us to set apart what is useful information
concerning the location of the QCP from finite size effects. 
For all the models here investigated, we did know their complete 
thermodynamic limit behavior as well as their quantum phase diagram.
As such, we were able to clearly show that we can 
obtain fairly good estimates for the location of the QCPs
for those models working with spin chains containing about 10 
qubits. But most importantly, this comparison allowed us to
identify several finite size effects that permeate all the 
models studied here. If those effects are properly taken into
account, we believe that it will be possible to correctly identify
a QCP using small systems even when we do not have access to its thermodynamic limit solution.

The second point of view, as anticipated in the last sentence above, assumes that we do not know much about 
the thermodynamic limit behavior of the system. 
This may happen because we either know the 
model (Hamiltonian) being studied but its thermodynamic limit 
solution is yet unknown or we do not even know the correct model
that we need to describe a given spin chain. In this situation, 
and in accord with the general finite size trends observed for 
all the models here investigated, we should implement the 
following checks or steps to distinguish a QCP from false
alarms coming from finite size effects. 

\textit{Step 1.} Whenever we work with small spin chains, 
it is probable that several abrupt changes in the fidelity 
$\overline{\mathcal{F}}_{ext}$
or in the trace distance 
$\overline{\mathcal{D}}_{int}$ will occur. This step-like behavior
may or may not indicate a QCP. In order to distinguish if an 
abrupt change in either $\overline{\mathcal{F}}_{ext}$ or 
$\overline{\mathcal{D}}_{int}$ is related to a QCP or to finite size effects, we should look at the high
temperature data. For all the models here presented, the finite
size step-like behavior disappears when we increase $T$. The 
remaining abrupt changes, although smoothed at high $T$, were all 
related to QCPs.

\textit{Step 2.} Compute, or measure, the curves of
$\overline{\mathcal{F}}_{ext}$ and $\overline{\mathcal{D}}_{int}$
for several spin chain sizes $L$ 
as functions of the tuning parameter.
Whenever those curves start to accumulate on top of each other, differing very little from 
the preceding smaller chain, we are probably close to a very good approximation for the thermodynamic limit.

\textit{Step 3.} After steps 1 and 2, we may still 
find spurious cusps or kinks in $\overline{\mathcal{F}}_{ext}$ 
and $\overline{\mathcal{D}}_{int}$ that are not 
associated with QCPs. Those spurious discontinuities in the 
derivatives of $\overline{\mathcal{F}}_{ext}$ and 
$\overline{\mathcal{D}}_{int}$ are almost always related to
the optimization over the sets $S_k$ of 
unitary operations available to Bob in order to get those
two quantities. Moreover, most of the time the cusps related to 
real QCPs do not coincide with the cusps marking the spots
where the set $S_k$ optimizing 
$\overline{\mathcal{F}}_{ext}$ or $\overline{\mathcal{D}}_{int}$ changes.

\textit{Step 4.} Cusps or inflections points that are actually
related to QCPs are always located in the same place, whether 
we extract them from $\overline{\mathcal{F}}_{ext}$ or from
$\overline{\mathcal{D}}_{int}$. The other cusps that are 
exclusively related to the optimization over the sets $S_k$ 
are not located in the same places in the curves of 
$\overline{\mathcal{F}}_{ext}$ and $\overline{\mathcal{D}}_{int}$.
This is a very useful strategy in discriminating false alarms and 
it highlights the importance of using both the external and the internal teleportation based QCP detectors to correctly pinpoint
the location of a QCP. Note that this strategy is also useful
in the thermodynamic limit since cusps related to the 
optimization over the sets $S_k$ also occur there.

We finish this section by calling attention to two other 
important features of the teleportation based QCP detectors that were fully discussed 
elsewhere.\cite{pav23,pav23b,pav24} The first one is related to
the behavior of $\overline{\mathcal{F}}_{ext}$ and 
$\overline{\mathcal{D}}_{int}$ as a function of the tuning 
parameter for each one of the models studied here. Looking at
Figs. \ref{fig3} and \ref{fig4}, at the top panels of 
Figs. \ref{fig6}, \ref{fig7}, and \ref{fig8}, and at 
Fig. \ref{fig9}, we realize that those curves are unique to each
model. Therefore, not only are we able to extract the QCPs 
from those curves but we are also able to discover the 
underlying model associated with those QCPs. In other words,
a given model has its own fingerprint, which is manifested in 
the unique behavior of $\overline{\mathcal{F}}_{ext}$ and 
$\overline{\mathcal{D}}_{int}$ for that particular model.

The second feature is related to the experimental facet of 
the teleportation based QCP detectors. In the standard way 
of experimentally studying QPTs, we need to measure the 
one- and two-point correlation functions to detect a 
QCP. The non-analytic properties of 
the one- and two-point correlation functions or of their 
derivatives suggest a QPT. The experimental implementation of
the teleportation based QCP detectors, though, only requires 
the measurement of one-point correlation
functions, before starting the teleportation protocol and 
after finishing it.\cite{pav23,pav23b,pav24} 
Furthermore, several platforms already realize tens to hundreds
of qubits that can be manipulated coherently for sufficiently 
long time to implement a quantum teleportation 
protocol.\cite{ron15}$\,^-\,$\cite{pet22} Therefore,
the experimental implementation of the present ideas are 
within reach using state of the art techniques that are 
already available.

\section{Conclusion}
\label{conclusion}

We studied the efficiency of the external and internal 
quantum teleportation based quantum critical point (QCP) detectors\cite{pav23,pav23b,pav24}
to correctly pinpoint the location of the QCPs for several different models when we only have
access to small spin chains. Moreover, we investigated 
the efficacy of those QCP detectors to properly extract 
the location of the QCP when only finite temperature data
are available. The simultaneous occurrence of these two constraints, namely, non zero temperature and small chains,
can be considered the ``worst case'' scenario. Indeed, 
the characterization of quantum phase transitions (QPTs) and 
the determination of where those transitions occur, i.e, the 
QCPs, are very challenging under those circumstances since 
a QPT is defined at $T=0$ and in the thermodynamic limit 
(infinite spin chains).\cite{sac99} As such, several strategies 
to overcome the finite size effects were given in order to
prevent ``false alarms'', i.e., an incorrect identification of 
a QPT that was actually a finite size effect or a ``glitch''
of the QCP detectors (see Sec. \ref{discussion}).

The models investigated here were one-dimensional 
spin-1/2 chains. Specifically, we studied the 
XXZ model with and without an external longitudinal 
magnetic field, the XX and XY models in an external transverse
magnetic field, and the Ising transverse model. We fixed our attention on spin chains containing at most 12 qubits and
in equilibrium with a thermal reservoir at temperature $T$.
We then numerically computed the canonical ensemble density 
matrix describing the whole chain, from which the relevant
quantities were extracted to apply the teleportation based 
QCP detectors.\cite{pav23,pav23b,pav24} 

For all the models investigated, we observed that for sufficiently
low temperatures very good estimates for the location of the QCPs
were obtained using spin chains with 10 or 12 qubits. The difference between the exact location of the QCPs and the location
extracted from those finite spin chains were always below a few 
percents. 
There was only one exception, related to the second QCP of the 
XXZ model in an external field. In this case, the spin chain with 12 qubits furnished the closest estimate to the correct location 
of the QCP, with a difference of about $10\%$ from the 
exact value of the QCP.  
However, fixing a given temperature, we noticed that
the greater $L$, the better the estimate of the QCP. 

We also noticed that for a fixed size $L$ of the spin chain,
the lower the temperature, the better the estimate of the QCP.
For most of the models and below a certain temperature threshold, 
the estimated QCPs tended to the exact thermodynamic limit 
values linearly as we decreased $T$. 
Thus, by extrapolating from the finite $T$ data
to $T=0$ led to very good predictions of the correct locations of those QCPs. 

Finally, we would like to call attention to the fact that all the models
investigated here were local, i.e., there were only nearest 
neighbor interactions. It would be interesting to investigate 
non-local models.\cite{sta17,ver17} We conjecture that the more non-local the model, the longer the spin chain must be in order to provide reasonable
estimates for the exact location of a QCP. Also, another 
interesting approach would be to test the efficiency of the 
teleportation based QCP detectors in disordered spin chains.
\cite{sta17,ver17,hoy06} How these tools 
to detect QCPs will work when we have disorder induced QPTs 
is an open problem so far.

\section*{Acknowledgements}

GR thanks the Brazilian agency CNPq (National
Council for Scientific and Technological Development)
for funding and CNPq/FAPERJ (State of Rio de Janeiro
Research Foundation) for financial support through the
National Institute of Science and Technology for Quantum Information. GAPR thanks the São Paulo Research
Foundation (FAPESP) for financial support through the
grant 2023/03947-0.


\end{document}